\documentclass[aps,prl,twocolumn,showpacs,floatfix]{revtex4}
\usepackage{graphics,graphicx}
\usepackage{color}
\usepackage{wrapfig}
\usepackage{enumerate}
\usepackage{amssymb,amsmath}
\usepackage{bm}
\usepackage{here}
\usepackage{ulem}

\twocolumngrid
\begin{document}
\unitlength = 1mm
\title{
  Magnetoelectric Behavior from $S=1/2$ Asymmetric Square Cupolas 
}

\author{
  Yasuyuki~Kato$^1$, Kenta~Kimura$^2$, 
  Atsushi~Miyake$^3$, Masashi~Tokunaga$^3$, Akira~Matsuo$^3$, Koichi~Kindo$^3$,
  Mitsuru~Akaki$^4$, Masayuki~Hagiwara$^4$, 
  Masakazu~Sera$^2$, Tsuyoshi~Kimura$^2$, and Yukitoshi~Motome$^1$}
\affiliation{
  $^1$Department of Applied Physics, University of Tokyo, Hongo, 7-3-1, Bunkyo, Tokyo 113-8656, Japan\\
  $^2$Division of Materials Physics, Graduate School of Engineering Science, Osaka University, Toyonaka, Osaka 560-8531, Japan\\
  $^3$Institute for Solid State Physics, The University of Tokyo, Kashiwa, Chiba 277-8581, Japan\\
  $^4$Center for Advanced High Magnetic Field Science, Graduate School of Science, Osaka University, Toyonaka, Osaka 560-0043, Japan
}

\date{\today}
%

\begin{abstract}
  Magnetoelectric properties are studied by a combined experimental and theoretical study of
  a quasi-two-dimensional
  material composed of square cupolas, Ba(TiO)Cu$_4$(PO$_4$)$_4$.
  The magnetization is measured 
  up to above the saturation field, and several anomalies are observed depending on the field directions.
  We propose a $S$=1/2 spin model with
  Dzyaloshinskii-Moriya interactions, 
  which well reproduces the full magnetization curves.
  Elaborating the phase diagram of the model,
  we show that the anomalies are explained by magnetoelectric phase transitions.
  Our theory also accounts for the scaling of the dielectric anomaly observed in experiments.
  The results elucidate the crucial role of the in-plane component of Dzyaloshinskii-Moriya
  interactions, which is induced by the noncoplanar buckling of square cupola.
  We also predict a `hidden' phase and another magnetoelectric response 
  both of which appear in nonzero magnetic field.
\end{abstract}
\maketitle

Spatial asymmetry is a source of interesting phenomena in a broad range of condensed matter physics. 
A well-known example is the molecular asymmetry of water H$_2$O, which leads to an electric polarization in each molecule. 
The asymmetry is at play also in magnets: the loss of inversion symmetry activates the asymmetric interactions through the relativistic spin-orbit coupling, such as the Dzyaloshinskii-Moriya (DM) interaction~\cite{dzyaloshinskii1958,moriya1960}. 
The asymmetric interactions lead to intriguing magnetism, e.g., weak ferromagnetism in antiferromagnets and spin-spiral ordering in helimagnets. 
They have also attracted growing interest as an origin of the magnetoelectric (ME) effect, that is, cross correlations between 
dielectricity and magnetism~\cite{cheong2007,tokura2014}.

\begin{figure}[ht]
  \centering
  \includegraphics[width=\columnwidth]{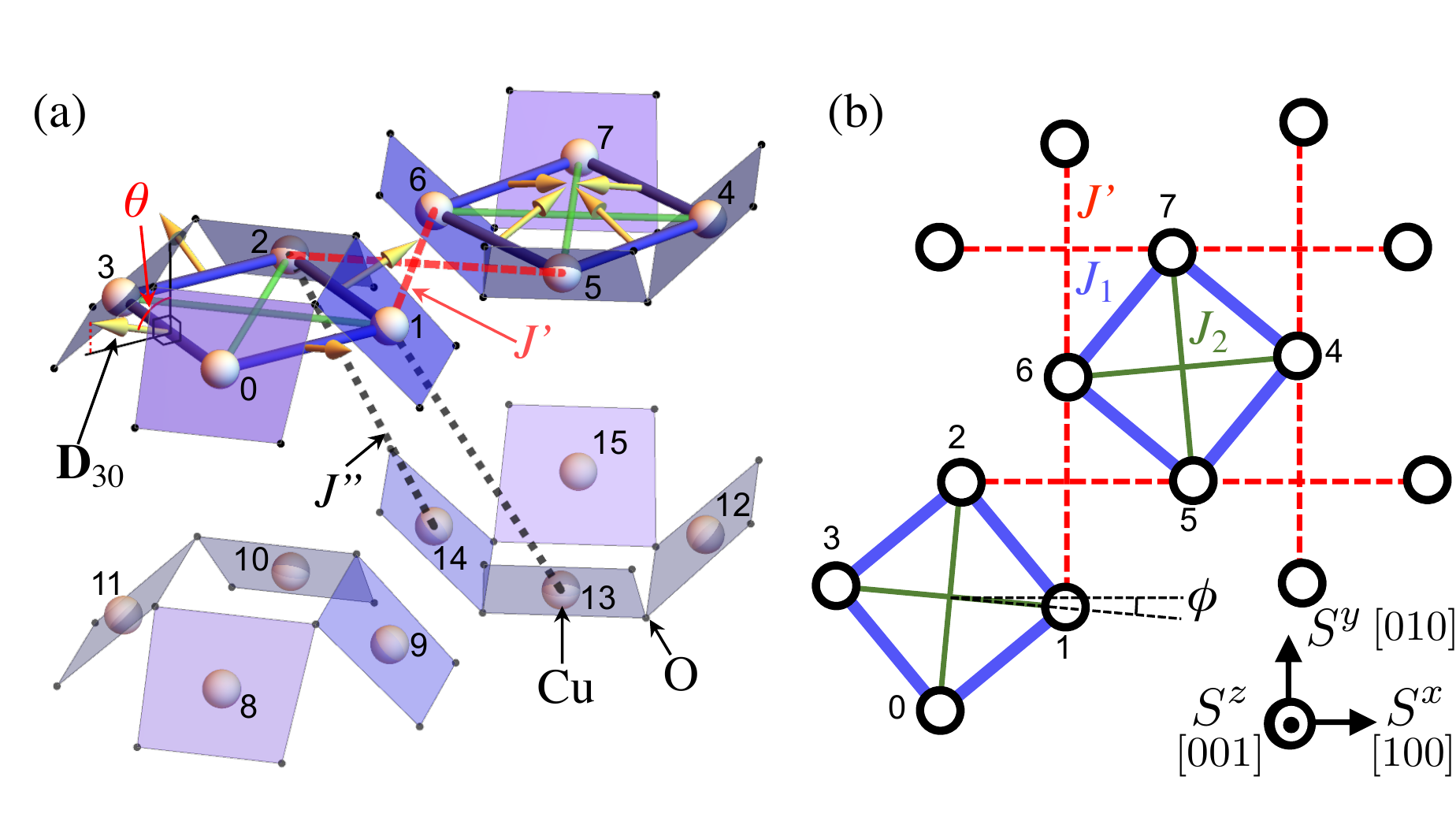}
  \caption{
    (a) Schematic of the 
    lattice structure, 
    which includes a pair of upward and downward square cupolas. 
    The spheres and black dots represent Cu cations and O ions, respectively.
    The yellow arrows on the blue bonds represent the DM vectors ${\bf D}_{ij}$ in the model in Eq.~(\ref{eq:model}).
    The intercupola couplings $J'$ and $J''$ are also shown.
    (b) Top view. 
    The blue,  green, and red dashed lines represent $J_1$, $J_2$, and $J'$ bonds in Eq.~\eqref{eq:model}, respectively.
  }
  \label{fig1}
\end{figure}

Recently, an interesting series of chiral antiferromagnets, 
$A$(TiO)Cu$_4$(PO$_4$)$_4$ ($A$ = Ba, Sr) with space group $P$42$_1$2, was newly synthesized~\cite{kimura2016a}. 
The materials have a quasi-two-dimensional structure, composed of an alternating array of
Cu$_4$O$_{12}$ clusters, as shown in Fig.~\ref{fig1}(a).
Each Cu$_4$O$_{12}$ cluster consists of four corner-sharing CuO$_4$ plaquettes, forming a noncoplanar buckled structure termed (irregular) square cupola. 
The asymmetric unit can carry ME-active magnetic multipoles~\cite{spaldin2008} associated with Cu spins. 
Indeed, a divergent anomaly of the dielectric constant was observed at
the N\'eel temperature ($T_{\rm N}$=9.5 K) in magnetic fields applied along the [100] and [1$\bar{1}$0] directions for $A$ = Ba~\cite{kimura2016}. 
Although the ME response was argued by the magnetic quadrupole associated 
with noncoplanar antiferromagnetic ordering, the microscopic understanding is not fully obtained. 
It is highly desired to clarify how the unique asymmetry arising from the square cupolas affects the magnetic and dielectric properties in this series of compounds.

In this Letter, combining experimental and theoretical studies, we clarify the microscopic mechanism of ME behavior in $A$(TiO)Cu$_4$(PO$_4$)$_4$.
First, from the magnetization measurement for the compound with $A$ = Ba up to above the saturation field,
we find several anomalies depending on the field direction. 
Then, considering the structual asymmetry, we construct a minimal theoretical model, which successfully reproduces the full magnetization curves within a cluster mean-field (CMF) approximation.
Elaborating the ground-state and finite-temperature ($T$) 
phase diagrams of the model, 
we show that the anomalies originate from ME phase transitions where the in-plane component of the DM vector plays a key role. 
Furthermore, we successfully reproduce the dielectric anomaly and its scaling behavior observed in the prior experiments~\cite{kimura2016}.
Our results indicate that the series of the compounds provides a unique playground where the
electric polarization is controllable not only by a magnetic field or $T$ but also 
by the asymmetry of square cupolas that modulates the orientation of DM vectors.

Single crystals of Ba(TiO)Cu$_4$(PO$_4$)$_4$ were grown by the flux method as described previously~\cite{kimura2016a}. 
Powder X-ray diffraction (XRD) measurements on crushed single crystals confirmed a single phase. 
For magnetization measurements, the crystals were oriented using Laue XRD. 
High-field magnetization in magnetic fields up to 69~T was measured at 1.4~K 
using an induction method with a multilayer pulsed magnet installed 
at the International MegaGauss Science Laboratory of Institute for Solid State Physics at The University of Tokyo. 
Multi-frequency electron spin resonance (ESR) measurements (600--1400 GHz) in pulsed magnetic fields were performed to obtain the $g$-values for the field directions along $[100]$, $[110]$ and $[001]$. 
The $g$-values were found to be isotropic within the experimental accuracy: $g \sim 2.20(5)$ for the three directions.

\begin{figure}[ht]
  \centering
  \includegraphics[width=0.9\columnwidth]{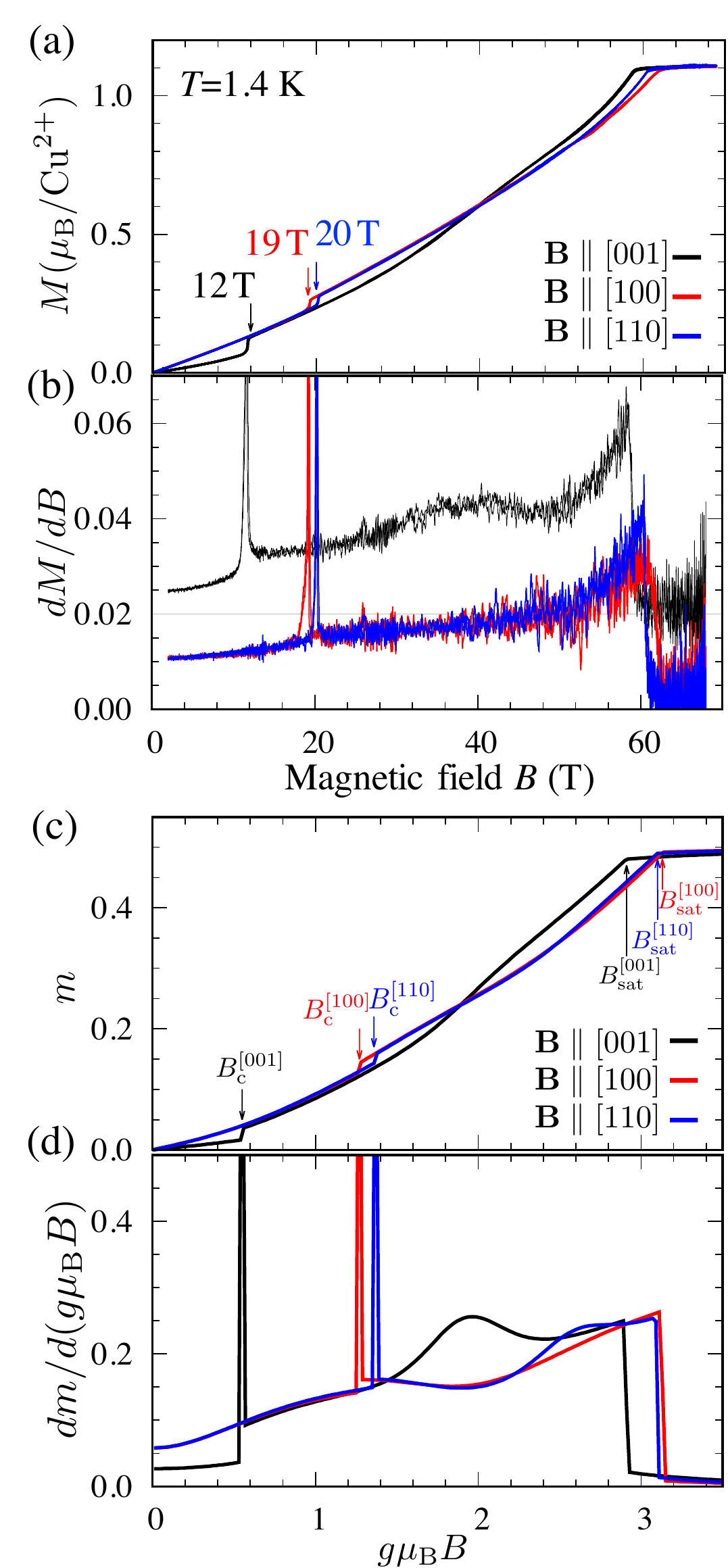}
  \caption{
    Magnetic field dependences of the magnetization and its field derivative
    in (a,b) experiment for Ba(TiO)Cu$_4$(PO$_4$)$_4$ at $T$=1.4~K
    ($dM/dB$ for ${\bf B}\parallel[001]$ is shifted by 0.02 for visibility) 
    and (c,d) theory for the model~\eqref{eq:model} in the ground state with 
    $J_1$=1, $J_2$=1/6, $J'$=1/2,  $J''$=1/100, $\theta$=$80^\circ$, and $D$=0.7. 
  }
  \label{fig2}
\end{figure}

Figure~\ref{fig2}(a) shows the experimental results of full magnetization curves at $T$=1.4~K $< T_{\rm N}$ for the magnetic field (${\bf B}$) 
applied along the $[100]$, $[110]$, and $[001]$ directions. 
In the low field region ($B=|{\bf B}|<12$~T), the slope of the magnetization $M$ is smaller for the out-of-plane field (${\bf B} \parallel [001]$) 
than for in-plane fields (${\bf B} \parallel [100]$ and $[110]$), which is consistent with the previous report~\cite{kimura2016}. 
On increasing $B$, we find a jump-like anomaly in $M$ for all the $B$ directions, 
whose critical field is $\sim 12$, $19$, and $20$~T for ${\bf B}\parallel[001]$, $[100]$, and $[110]$, respectively;
the anomalies are more clearly seen in the field derivative $dM/dB$ in Fig.~\ref{fig2}(b). 
On further increasing $B$ above $B \sim  60$~T, 
the magnetization for all the $B$ directions shows a saturation at $\sim$1.1~$\mu_{\rm B}/{\rm Cu}^{2+}$. 
The saturation-magnetization values are corrected by the $g$-values determined by the ESR. 
We note that $dM/dB$ shows a hump at $B \sim 40$~T only for ${\bf B}\parallel[001]$ 
as shown in Fig.~\ref{fig2}(b).

For understanding of the peculiar magnetization curves, 
we consider a localized spin model associated with $S$=1/2 spin degrees of freedom of each Cu cation.
We take into account four dominant exchange interactions~\cite{kimura2016}: 
the intracupola exchange interactions $J_1$ and $J_2$, together with the two intercupola couplings, 
intralayer $J'$ and interlayer $J''$ (see Fig.~\ref{fig1}). 
Besides, considering the loss of inversion symmetry at the centers of $J_1$ bonds, we take into account the DM interaction originating from the spin-orbit coupling.
The Hamiltonian reads
\begin{eqnarray}
  \mathcal{H}
  =\sum_{\langle i,j \rangle} \left[
    J_1 {\bf S}_i \cdot {\bf S}_j
    - {\bf D}_{ij}
    \cdot {\bf S}_i \times {\bf S}_j 
    \right]
+ J_2 \sum_{\langle\langle i,j \rangle\rangle} {\bf S}_i \cdot {\bf S}_j \nonumber\\
+ J' \sum_{(i,j)} {\bf S}_i \cdot {\bf S}_j
+ J'' \sum_{((i,j))} {\bf S}_i \cdot {\bf S}_j
-g\mu_{\rm B}\sum_i {\bf B}\cdot{\bf S}_i,
\label{eq:model}
\end{eqnarray}
where ${\bf S}_i$ represents $S$=1/2 spin at 
site $i$.
The sums for $\langle i,j \rangle$, $\langle\langle i,j \rangle\rangle$,
$(i,j)$, and $((i,j))$ run over $J_1$, $J_2$, $J'$, and $J''$ bonds, respectively.
For the intracupola exchange coupling constants, we adopt the estimates from the {\it ab initio} calculation:
$J_1$=3.0~meV and $J_2$=0.5~meV~\cite{kimura2016}; we set $J_1$ as the unit of energy, namely, $J_1$=1 and $J_2$=1/6. 
On the other hand, for the intercupola $J'$, 
we set a larger value $J'$=1/2 than the {\it ab initio} estimate $J'$=0.7~meV, 
because small $J' \lesssim 0.4$
leads to a nonmagnetic singlet state in the CMF approach (see below);
we set $J''$=1/100.
The last term in Eq.~(\ref{eq:model}) represents the Zeeman coupling
where $g$ and $\mu_{\rm B}$ are the isotropic $g$-factor and the Bohr magneton, respectively.

In the second term in Eq.~\eqref{eq:model},
referring the Moriya rules~\cite{moriya1960}, 
we take the DM vector ${\bf D}_{ij}$ in the plain perpendicular to the $J_1$ bond 
connecting $i$ and $j$ sites with the angle $\theta_{ij}$ from the $[001]$ axis
[see the yellow arrows in Fig.~\ref{fig1}(a)].
Note that the asymmetric buckling of CuO$_4$ plaquettes in the square cupola induces the 
in-plane component of ${\bf D}_{ij}$.
The sign of ${\bf D}_{ij}$ is reversed between the upward and downward cupolas from the symmetry.
We assume uniform $\theta=\theta_{ij}$ and $D \equiv |{\bf D}_{ij}|$, and tune the values of $\theta$ and $D$
so as to reproduce the magnetization curve in experiments.
While the system has the chirality from the alternating twist of square cupolas, $\phi$ in Fig.~\ref{fig1}(b), 
we omit it for simplicity; 
the following results remain intact for small $\phi$ in the real compounds~\cite{kimura2016a}.

We calculate the magnetic properties of the model~\eqref{eq:model} by a CMF method 
in which the intracupola interactions
$J_1$ and $J_2$ are dealt with by the exact diagonalization, while the intercupola interactions $J'$ and $J''$ 
are decoupled by the mean fields as
${\bf S}_i \cdot {\bf S}_j \simeq  \langle {\bf S}_i \rangle \cdot {\bf S}_j 
+ {\bf S}_i \cdot \langle {\bf S}_j \rangle - \langle{\bf S}_i \rangle \cdot \langle{\bf S}_j \rangle $.
In the CMF calculations, we assume the 16-sublattice magnetic unit cell shown in Fig.~\ref{fig1}(a).

Figures~\ref{fig2}(c) and \ref{fig2}(d) show the $B$ profiles of magnetization per site $m$
and its field derivative $dm/d(g\mu_{\rm B}B)$, respectively, 
obtained by the CMF calculations with $\theta$=$80^\circ$ and $D$=0.7 
(we will discuss how to optimize the parameters later). 
The results well reproduce the experimental data in Figs.~\ref{fig2}(a) and \ref{fig2}(b) in the following points. 
(i) $m$ shows a jump-like anomaly, whose magnetic field depends on the field direction: 
the anomaly for the out-of-plane field (${\bf B} \parallel [001]$)
appears at $B_{\rm c}^{[001]}$, which is considerably smaller 
than $B_{\rm c}^{[100]}$ and $B_{\rm c}^{[110]}$ for the in-plane fields
${\bf B} \parallel [100]$ and $[110]$, 
respectively, and $B_{\rm c}^{[100]}$ is slightly lower than $B_{\rm c}^{[110]}$. 
(ii) 
  The saturation fields, $B^{[100]}_{\rm sat}$, $B^{[110]}_{\rm sat}$, and $B^{[001]}_{\rm sat}$
  for ${\bf B} \parallel[100]$, $[110]$, and $[001]$, respectively,
  satisfy the relation, $B^{[100]}_{\rm sat} > B^{[110]}_{\rm sat} > B^{[001]}_{\rm sat}$.
(iii) 
$dm/d(g\mu_{\rm B}B)$ is smaller for ${\bf B} \parallel [001]$ 
than for ${\bf B} \parallel [100]$ and $[110]$ in the low field region. 
(iv)
$dm/d(g\mu_{\rm B}B)$ exhibits a hump at an intermediate field for ${\bf B}\parallel [001]$.

\begin{figure*}[ht]
  \centering
  \includegraphics[width=\textwidth]{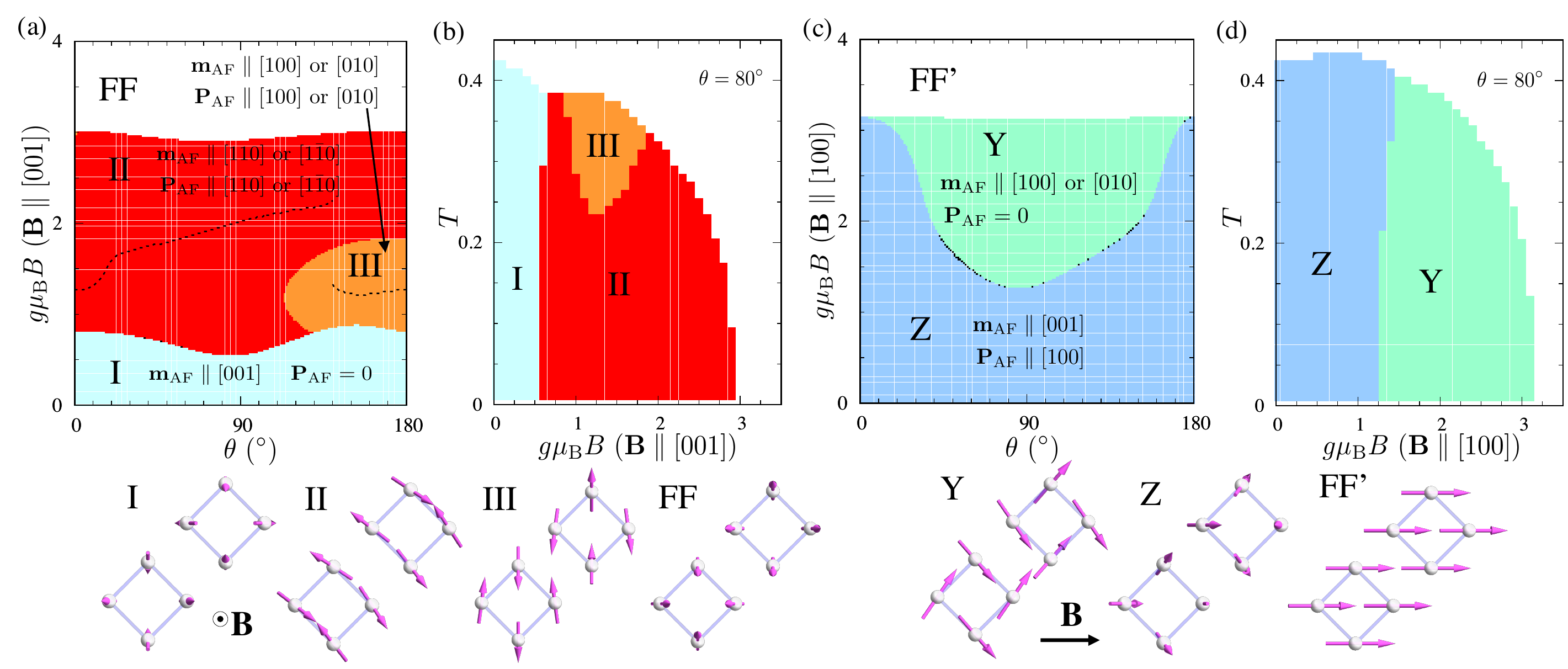}
  \caption{    
    Ground-state and finite-$T$ phase diagrams 
    with the magnetic fields (a,b) ${\bf B}\parallel[001]$
    and (c,d) ${\bf B}\parallel[100]$.
    In the ground-state phase diagrams (a,c), the angle $\theta$ between ${\bf D}_{ij}$ and $c$-axis 
    is varied, and the other parameters are set to be the same as Figs.~\ref{fig2}(c,d).
    In the finite-$T$ 
    phase diagrams (b,d), we set $\theta$=$80^\circ$ as in Figs.~\ref{fig2}(c,d).
    The labels I, II, III, Z, and Y identify different antiferromagnetic ordered phases where ${\bf m}_{\rm AF} \neq 0$.
    FF and FF' are for the forced ferromagnetic phases above the saturation fields.
    The dashed line in (a) represents the peak position of the hump in $dm/dB$. 
    Representative spin configurations in each phase are also shown~\cite{supp}.
  }
  \label{fig3}
\end{figure*}
\begin{figure}[ht]
  \centering
  \includegraphics[width=\columnwidth]{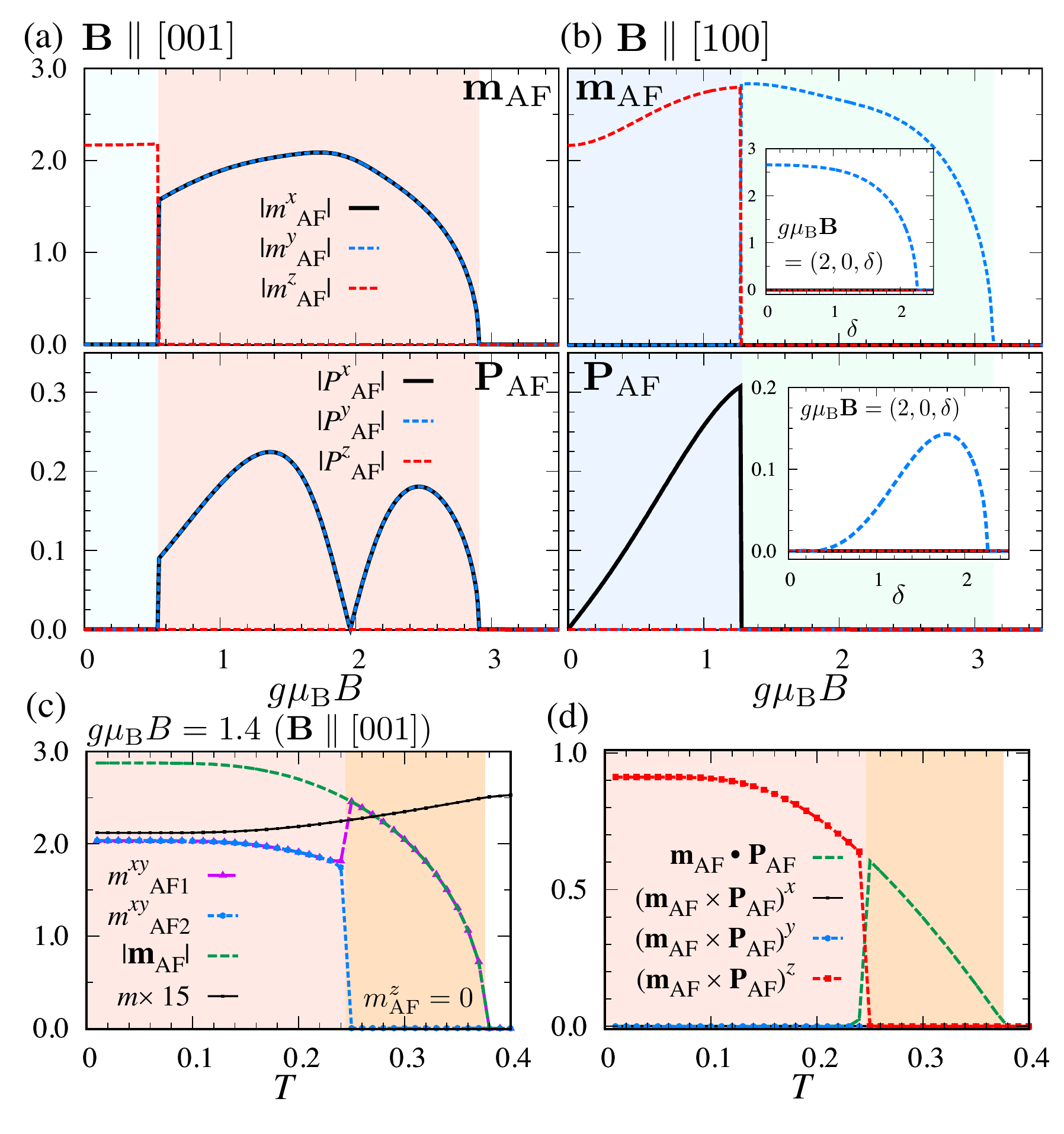}
  \caption{    
    $B$ and $T$ dependences of the antiferromagnetic order parameter ${\bf m}_{\rm AF}$ and 
    the interlayer-antiferroic electric polarization
    $ {\bf P}_{\rm AF}$ 
    with the parameter set used in Figs.~\ref{fig2}(c,d).
    The $B$ dependences are computed at $T$=0 for (a) ${\bf B}\parallel[001]$ and (b)  ${\bf B}\parallel[100]$,
    while the $T$ dependences (c,d) are computed at $B$=1.4 (${\bf B}\parallel [001]$).
    In addition to ${\bf m}_{\rm AF}$, the magnetization $m$ is also plotted in (c);
      $m^{xy}_{\rm AF1} = \max[ m^x_{\rm AF}, m^y_{\rm AF}]$ and
      $m^{xy}_{\rm AF2} = \min[ m^x_{\rm AF}, m^y_{\rm AF}]$.
    The inner and outer products of ${\bf m}_{\rm AF}$ and $ {\bf P}_{\rm AF}$ are plotted in (d).
    The insets of (b) show ${\bf m}_{\rm AF}$ and $ {\bf P}_{\rm AF}$ 
    for ${\bf B}=(2,0,\delta)$ with varying $\delta$.
  }
  \label{fig4}
\end{figure}

We find that the jump-like anomalies in $m$ originate from the ME phase transitions by elaborating the phase diagram of the model in Eq.~(\ref{eq:model}). 
Figures~\ref{fig3}(a) and \ref{fig3}(c) show the ground-state phase diagrams
while varying $\theta$ and $B$ for ${\bf B}\parallel [001]$ and $[100]$, respectively, at $D$=0.7.
To identify each phase, we compute
the anitiferromagnetic order parameter ${\bf m}_{\rm AF}$ and 
the interlayer-antiferroic electric polarization
$ {\bf P}_{\rm AF}$, which are defined as
${\bf m}_{\rm AF} = \sum_i (-1)^i p_i {\bf S}_i$ 
and 
$ {\bf P}_{\rm AF} = \sum p_i q_i {\bf n}_{ij} \langle {\bf S}_i \cdot {\bf S}_j \rangle$, 
respectively.
The sums run over all the 16 sites in the magnetic unit cell shown in Fig.~\ref{fig1}(a), 
$p_i=+1$($-1$) for the upper (lower) layer, and $q_i=+1$($-1$) for the upward(downward) cupolas; 
${\bf n}_{ij}$ is the normalized vector pointing from the center of $ij$ bond to the oxygen site shared by the sites $i$ and $j$.
We assume that the electric polarization arises from the 
exchange striction mechanism~\cite{sergienko2006,picozzi2007,delaney2009};
we confirmed that the inverse DM mechanism~\cite{katsura2005}
leads to qualitatively similar behavior~\cite{supp}.

At zero field, the ground state of our model exhibits a 
noncoplanar antiferromagnetic order~\cite{supp}, similar to that found in the neutron scattering experiment~\cite{kimura2016}.
For ${\bf B}\parallel [001]$, we find three phases below the saturation field, as shown in Fig.~\ref{fig3}(a).
The phases I, II, and III are antiferromagnetically ordered phases; 
${\bf m}_{\rm AF} \parallel {\bf B}$ and $ {\bf P}_{\rm AF}$=0 in the phase I, 
while ${\bf m}_{\rm AF}\perp {\bf B}$ and $ {\bf P}_{\rm AF} \neq 0$ in II and III [see also Fig.~\ref{fig4}(a)].
Thus, the phases II and III have 
an interlayer-antiferroic polarization induced by the antiferromagnetic ordering. 
The magnetization jump appears at the boundary between the paraelectric phase I and the antiferroelectric phase II [$B_{\rm c}^{[001]}$ in Fig.~\ref{fig2}(c)]. 
We note that the hump appears in $dm/dB$ in the phase II, whose position increases as increasing $\theta$ [the dashed line in Fig.~\ref{fig3}(a)].
This hump originates from quantum fluctuations within the clusters and the in-plane component of ${\bf D}_{ij}$.
For ${\bf B}\parallel [100]$, on the other hand, 
we find two antiferromagnetic phases, Z and Y, in both of which
${\bf m}_{\rm AF} \perp {\bf B}$; ${\bf m}_{\rm AF} \parallel [001]$ and $ {\bf P}_{\rm AF} \neq 0$ in Z, while ${\bf m}_{\rm AF} \parallel [010]$ and $ {\bf P}_{\rm AF}$= 0 in Y, as shown in Fig.~\ref{fig3}(c) [see also Fig.~\ref{fig4}(b)].
In this case also, the magnetization jump appears at the antiferro-paraelectric phase boundary between Z and Y [$B_{\rm c}^{[100]}$ in Fig.~\ref{fig2}(c)].

Let us comment on how we choose $\theta$ and $D$ used above.
With regard to $D$, we find that a sufficiently large $D\sim 0.7$ is necessary to reproduce the experimental value of the ratio 
$B_{\rm c}^{[001]}/B_{\rm sat}^{[001]}\sim 1/5$~\cite{supp}.
Similar large $D$ has been reported for other Cu-based magnets~\cite{yang2012,sannigrahi2015}. 
For $\theta$, we find that $B_{\rm c}^{[100]}$ takes minimum at $\theta \simeq 80^\circ$, as shown in Fig.~\ref{fig3}(c). 
Optimizing the values of $D$ and $\theta$ to make the ratios $B_{\rm c}^{[001]}/B_{\rm sat}^{[001]}$ and 
$B_{\rm c}^{[100]}/B_{\rm sat}^{[100]}$ closest to the experimental estimates, we obtain $D$=0.7 and $\theta$=$80^\circ$. 
We note that the value of $\theta$ is in between the weak and strong crystal field limits~\cite{supp}.
The in-plane component of ${\bf D}_{ij}$ induced by nonzero $\theta$ is crucial to reproduce the experiments.

Figures~\ref{fig3}(b) and \ref{fig3}(d) show the finite-$T$ phase diagrams at $\theta$=$80^\circ$ and $D$=0.7 with ${\bf B}\parallel[100]$ and $[001]$, respectively.
Interestingly, the phase III appears only at finite $T$ and finite $B$, which evolves from that in the large $\theta$ region at zero $T$ in Fig.~\ref{fig3}(a). 
While $m$ shows a jump at the transition between the phases I and II,
a singularity is hardly discernible between II and III [see Fig.~\ref{fig4}(c)].
Thus, the phase III is a `hidden' phase in the magnetization measurement. 
Nonetheless, at the II-III phase boundary,
$ {\bf P}_{\rm AF}$ changes the direction from $[110]$ to $[100]$,
and simultaneously, ${\bf m}_{\rm AF}$ is switched from 
${\bf m}_{\rm AF} \perp  {\bf P}_{\rm AF}$ to ${\bf m}_{\rm AF} \parallel  {\bf P}_{\rm AF}$ [see Fig.~\ref{fig4}(d)].
Thus, the `hidden' phase III will be identified by the ME measurement in future.

\begin{figure}[ht]
  \centering
  \includegraphics[width=\columnwidth]{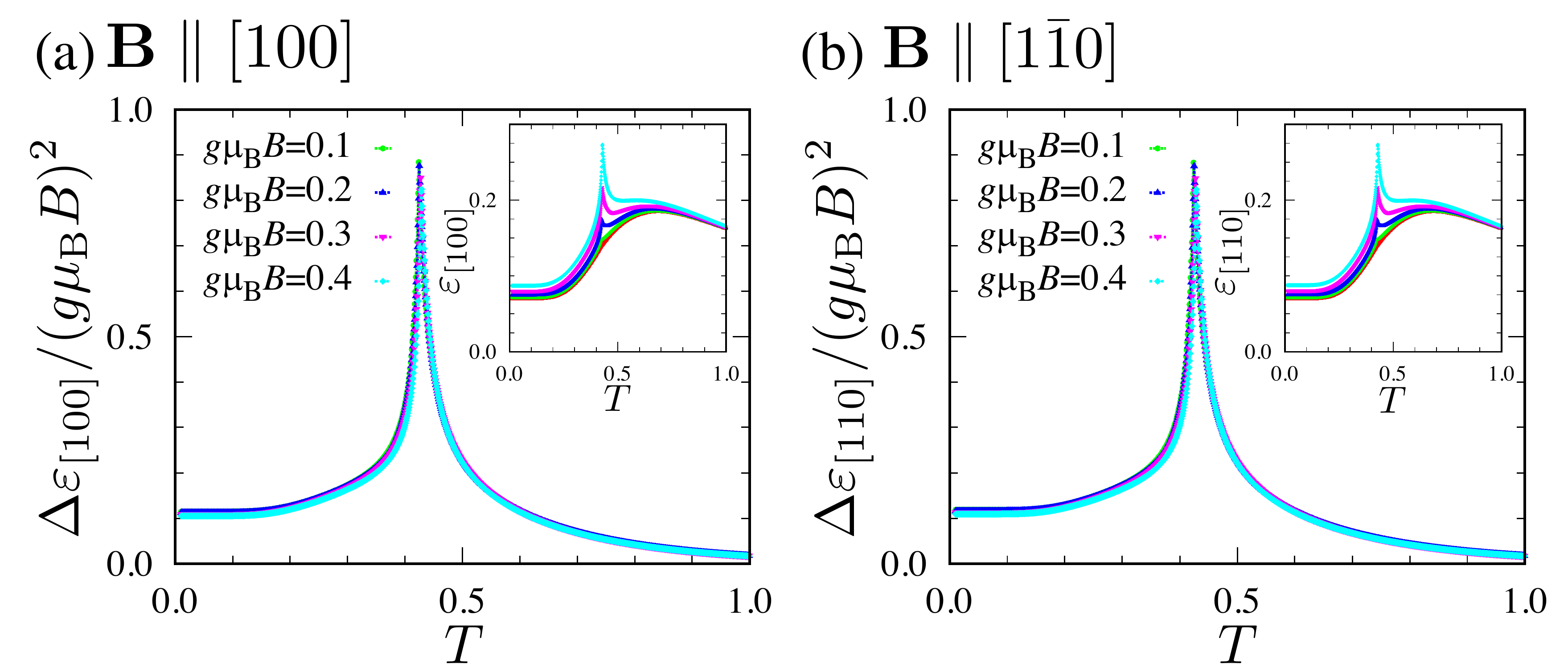}
  \caption{
    The scaling plot for the field-induced component of the dielectric constants:
      (a)~$\Delta \varepsilon_{[100]}$ and (b)~$\Delta \varepsilon_{[110]}$
      with ${\bf B}\parallel [100]$ and ${\bf B}\parallel [1\bar{1}0]$, respectively.
      The insets show 
      (a) $\varepsilon_{[100]}(B)$ and (b) $\varepsilon_{[110]}(B)$.
  }
  \label{fig6}
\end{figure}

Finally, let us discuss the ME behavior at finite $T$ in the low field region.
In the previous experiment,
the dielectric constants $\varepsilon_{[100]}$
and $\varepsilon_{[110]}$ show an anomaly at the antiferromagnetic transition in ${\bf B} \parallel  [100]$ and
$[1\bar{1}0]$, respectively, but not in ${\bf B} \parallel [001]$; furthermore, the field-induced components scale with $B^2$~\cite{kimura2016}. 
We here compute the dielectric constant as 
$\varepsilon_{[{\rm abc}]}=(\langle  {\bf P}\cdot {\bf n} \rangle_{{\bf E}=\Delta E {\bf n}} - \langle  {\bf P} \cdot {\bf n} \rangle_{{\bf E}=0})/\Delta E$,
where ${\bf n}$ is the normalized vector directing (a,b,c), for the Hamiltonian $\mathcal{H}- {\bf E} \cdot  {\bf P}$;
${\bf E}$ is the electric field, and $ {\bf P}$ is the net polarization defined as 
$ {\bf P} \equiv \sum_{\langle i,j \rangle} q_i {\bf n}_{ij} \langle {\bf S}_i \cdot {\bf S}_j \rangle$, (we set $\Delta E$=0.01).
As shown in the insets of Fig.~\ref{fig6},
both $\varepsilon_{[100]}$ and $\varepsilon_{[110]}$ show a cusp at the antiferromagnetic transition
as observed in experiments, corresponding to the onset of the antiferroic polarization in the phase Z. 
Moreover, our results successfully reproduce the scaling behavior $\Delta\varepsilon = \varepsilon(B) - \varepsilon(0) \propto B^2$ for both $\varepsilon_{[100]}$ and $\varepsilon_{[110]}$, as shown in Fig.~\ref{fig6}.
This scaling is naturally explained by the $B$-linear behavior of the in-plane polarization 
[see the lower panel of Fig.~\ref{fig4}(b)]
and $\varepsilon_{[{\rm abc}]} \sim \langle ( {\bf P}\cdot {\bf n})^2 \rangle$ expected from the fluctuation-dissipation theorem.

Our exhaustive analysis predicts a ME  response also in the phase Y: 
$ {\bf P}_{\rm AF} \parallel [010]$ is induced by applying ${\bf B} \parallel [001]$ in addition to ${\bf B} \parallel [100]$ [see the insets of Fig.~\ref{fig4}(b)]. 
This is associated with the ME coupling $\propto B^{[001]} E^{[010]}$ in the free energy~\cite{supp}. 
It will be interesting to confirm the ME response in future experiments.

  In summary, we have clarified the ME phase diagram of the antiferromagnets composed of square cupolas by a combined experimental and theoretical study.
  Our results elucidated how the asymmetric units made of small number of spins activate the ME responses through antisymmetric interactions arising from the spin-orbit coupling.
  The present study will provide a hint for the design of new type of ME-active materials originating from such asymmetric units.

\begin{acknowledgments}
The authors thank K.~Yamauchi for fruitful discussions.
This work was supported by JSPS Grant Numbers 24244059, 25220803, 26800199, 15K05176,
24244058, 16K05413, and 16K05449, and by a research grant from The Murata Science Foundation. 
Numerical calculations were conducted on the supercomputer system in ISSP, The University of Tokyo.

\end{acknowledgments}
\bibliographystyle{apsrev}
\bibliography{cu4}

\end{document}